\documentclass[12pt]{iopart}
\newcommand{\R}{{\mathbb{R}}}

\usepackage{epsfig,amsfonts,amscd}

\begin{document}
\title{ Furcation of resonance sets for one-point interactions
 }
\author{A.V. Zolotaryuk}
\address
{Bogolyubov Institute for Theoretical Physics, National Academy of
Sciences of Ukraine, Kyiv 03680, Ukraine}

\date{\today}

\begin{abstract}
Families of one-point interactions are derived from the system consisting of 
regularized two- and 
three-delta potentials using different paths of the convergence of  
corresponding transmission matrices in the squeezing limit. This limit is
controlled by the relative rate of shrinking the width of 
 delta-like functions and the distance between these functions using the power
parameterization: width $l =\varepsilon^{\mu -1}$, $\mu \in [2,\, \infty]$
(for width) and $r = \varepsilon^\tau$, $\tau \in [1,\, \infty]$ (for distance).
It is shown that at some values of real coefficients (intensities $a_1$, $a_2$ and $a_3$) 
at the delta potentials, the transmission across
the limit point interactions is non-zero, whereas outside these (resonance) values the one-point 
interactions are  opaque splitting the system at the  point of singularity into  two 
independent subsystems.  The resonance sets of intensities at which a non-zero 
transmission occurs are proved to be of four types depending on the way of squeezing
the regularized system to one point. In its turn, 
on these sets the limit one-point interactions
are observed to be either single- or multiple-resonant-tunnelling potentials
also depending on the squeezing way. In the two-delta case the resonance sets 
are curves on the $(a_1,a_2)$-plane and surfaces in the  $(a_1,a_2,a_3)$-space
for the three-delta system. A new phenomenon of furcation of  
single-valued resonance sets to multi-valued ones
is observed under approaching the parameter $\mu >2$ to the value $\mu =2$.
  \end{abstract}



\section{Introduction}

The models described by the Schr\"{o}dinger operators with singular zero-range potentials 
have widely been discussed in both the physical and mathematical literature
(see books~\cite{do1,do2,a-h,ak} for details and references). 
 These models admit exact closed analytical solutions
which describe realistic situations using different approximations via
Hamiltonians describing point interactions \cite{pc,bft,enz,c-g,ca,em}.
 Currently, because of the rapid progress in
fabricating nanoscale quantum devices, of particular importance is the point modelling of
different structures like quantum waveguides 
\cite{acf,ce}, spectral filters \cite{tc1,tc2} or infinitesimally thin sheets
 \cite{z,zz1,zz2}. A  whole body of literature (see, e.g., 
\cite{s,gh87,g,k,adk,cnp,cnt,an,n1,n2,gnn,acg,l1}, a few
to mention), including the very recent studies \cite{bn,gggm,l2,z15,kp,djp,gmmn,knt} 
with references therein, has been published  
where  the  one-dimensional  Schr\"{o}dinger operators with potentials given 
in the form of distributions are shown to 
exhibit a number of peculiar features with possible applications to quantum physics. 
A detailed list of references on this subject can also 
be found in the recent review \cite{km}.
On the other hand, using some particular regular approximations of the potential expressed
in the form of the derivative of Dirac's delta function, a number of  
interesting resonance properties of quantum particles
 tunnelling through this point potential  has been observed \cite{c-g,zci1,tn,zz14}. 
Particularly, it was found that at some values of the potential strength of the
$\delta'$-potential the transmission across this barrier is non-zero, whereas 
outside these values the barrier is fully opaque. In general terms, 
the existence of  such resonance sets in the space
of potential intensities has rigorously been established for a whole class of 
approximations of the derivative delta potential
by Golovaty with coworkers \cite{gm,gh,m1,g1,m2,gh1,g2}. This type of point interactions 
may be referred to as `resonant-tunnelling $\delta'$-potentials'. These results 
differ from those obtained within Kurasov’s theory \cite{k} which was 
developed for the distributions 
defined on the space of functions discontinuous at the point of singularity.
Here the limit point interaction is also called a $\delta'$-potential. 
The common feature of Kurasov's  point potential and a resonant-tunnelling 
$\delta'$-potential is that the
transmission matrices  of both these interactions are of the diagonal form, but
the elements of these matrices are different.  It is of
interest therefore to 
 to find a way where it would be possible to describe both these types in a 	aunique regularization scheme starting from the same initial regularized 
potential profile.

In the present work we address the problem on the relation between  the 
point interactions realized within Kurasov's theory and the resonant-tunnelling 
$\delta'$-potentials studied in \cite{c-g,zci1,tn,zz14,gm,gh,g1,gh1,g2,zpla10}.
Similarly to these papers, we explore
 the one-dimensional stationary Schr\"{o}dinger equation
\begin{equation}
-\, d^2\psi(x)/dx^2 + V_{\varepsilon }(x)\psi(x) =E\psi(x)
\label{1}
\end{equation} 
where $\psi(x)$ is the wavefunction and $E$ the energy of a particle.
The potential $V_{\varepsilon }(x)$ with a squeezing parameter $\varepsilon >0$
shrinks to one point, say $x =0$, as $\varepsilon \to 0$. One of the ways to realize
limit point interactions  is to choose the potential  
$V_{\varepsilon }(x)$  in the form of a sum 
of several  Dirac's delta functions as follows \cite{bn,cs,an1}
\begin{equation}
V_\varepsilon(x) = \sum_{j =1}^N c_j(\varepsilon) \delta(x - r_j(\varepsilon)),
~~r_j(\varepsilon ) \in \R,
\label{2}
\end{equation}
where all $r_j(\varepsilon)  \to 0$ and $c_j(\varepsilon) \to \pm \infty$ as $\varepsilon \to 0$.
The particular case of the three-delta spatially symmetric potential (\ref{2}), in the limit as
 the distances between the $\delta$-functions tend to zero, has been studied by
Cheon and Shigehara \cite{cs}, and Albeverio and Nizhnik \cite{an1}. In this limit 
 a whole four-parameter family of point interactions has been constructed,
independently on whether or not  potential (\ref{2}) has a distributional limit. 
Here we follow the approach developed   by Exner, Neidhardt and Zagrebnov
\cite{enz}, who have approximated the $\delta$-potentials  by regular functions and 
constructed a one-point limit interaction. In particular, they have proved that the  limit 
takes place if the distances between the `centers' of regularized potentials 
tend to zero sufficiently slow relatively to shrinking the $\delta$-like potentials. A similar 
research \cite{bft} concerns about 
 the convergence of regularized $\delta$-like structures to point potentials
in higher dimensions. 

In this paper we focus on the two cases when  potential  (\ref{2})
 consists of two ($N=2$) and three ($N=3$) $\delta$-potentials
separated equidistantly  by  a function $r(\varepsilon)$ 
that  tends to zero as $\varepsilon \to 0$. All the coefficients at the $\delta$-functions 
are specified as $c_j =- a_j/\varepsilon$ where $a_j$'s (`intensities', `charges' or
`amplitudes') are non-zero constants. The sign `-' has been chosen for convenience in the following
notations, so that negative values of $a_j$ correspond to a $\delta$-barrier and positive
ones to a $\delta$-well. Thus,  in the case with $N=3$ we have
\begin{equation}
\!\!\!\!\!\!\!\!\!\!\!\!\!\!\!\!\!\!\!
V_{\varepsilon r}(x) =- \varepsilon^{-1}\left[ a_1 \delta(x) + a_2\delta(x-r) + a_3\delta(x-2r)
\right], ~~(a_1,a_2,a_3) \in \R^3 \setminus \{0\}. 
\label{3}
\end{equation}
For the case of two $\delta$-potentials, we just set in (\ref{3}) $a_3= 0$, so that
$(a_1,a_2) \in \R^2 \setminus \{0\}$. 
The transmission matrices for the two- and three-delta potentials are the products
$\Lambda_{ \varepsilon r} =  \Lambda_2 \Lambda_0 \Lambda_1$ and 
$\Lambda_{ \varepsilon r} = \Lambda_3\Lambda_0 \Lambda_2 \Lambda_0 \Lambda_1$, 
respectively, where
\begin{eqnarray}
\!\!\!\!\!\!\!\!\!\!\!\!\!\!\!\!
  \Lambda_0  =  \left( \begin{array}{cc} ~\cos(kr)~~~~~ k^{-1}\sin(kr) \\
-\, k \sin(kr)  ~~~~~ \cos(kr) \end{array} \right) , ~~~~
\Lambda_j  =  \left( \begin{array}{cc} ~~~1~~~~ ~0 \\
-a_j/\varepsilon  ~ ~1 \end{array} \right) ,~~j=1,2,3.
\label{4}
\end{eqnarray}

We restrict ourselves to
 the most simple approximation of the $\delta$-potentials by 
piecewise constant functions resulting in a three (for $N=2$) and a  five (for $N=3$)
layered potential profile.
In the limit as both the width of $\delta$-like functions and the distance between them tends 
to zero simultaneously we obtain a family  of one-point interactions. 
 We observe that, starting from the same profile of the three- and
five-layered structure that approximates potential (\ref{2}), 
 the limit point interactions crucially depend on the relative rate of 
 tending the width of layers and the distance between them to zero.   Within this approach
one can realize both the  point interactions obtained 
within Kurasov's theory and the resonant-tunnelling potentials.

\section{A piecewise constant approximation of the $\delta$-potentials  }

Let us  approximate the $\delta$-potentials in (\ref{3}) by piecewise constant functions. Then
 potential (\ref{3}) is replaced by the rectangular function
\begin{equation}
\!\!\!\!\!\!\!\!\!\!\!\!\!\!\!\!\!\!\!\!\!\!\!\!\!\!\!\!\!
V_{\varepsilon lr}(x)= \left\{ \begin{array}{ll}
~~~ 0 &   \mbox{for}~~ -\infty < x < 0, ~ l < x < l+r , \\
& ~~~~~ ~2l +r <x< 2(l+r),~ 3l+2r < x< \infty , \\
- a_j / \varepsilon l ~ & \mbox{for}~~ (j-1)(l+r)< x < j(l+r)-r  ,~j=1,2,3,
\end{array} \right. 
\label{5}
\end{equation} 
and, as a result, all the matrices 
$\Lambda_j$, $j=1,2,3$, in the product for $\Lambda_{\varepsilon r}$ are replaced by 
\begin{eqnarray}
 \Lambda_{j,l}  =  \left( \begin{array}{cc} \cos(k_j l)~~ ~~~~k_j^{-1}\sin(k_jl) \\
-\, k_j \sin(k_j l)  ~~~~~ ~~~\cos(k_j l) \end{array} \right) ,
\label{6}
\end{eqnarray}
where 
\begin{equation}
 k_j : = \sqrt{  k^2 + a_j/\varepsilon l  }\, ,~~ k := \sqrt{E}\, , ~~j=1,2,3.
\label{7}
\end{equation}
In other words, the regularized
transmission matrix $\Lambda_{\varepsilon lr}$ defined by the relations
\begin{eqnarray}
\!\!\!\!\!\!\!\!\!\!\!\!\!\!\!\!\!\!\!\!
\left( \begin{array}{cc} \psi(x_2)  \\
\psi'(x_2) \end{array} \right) 
 = \Lambda_{\varepsilon lr} \left(
\begin{array}{cc} \psi(x_1)   \\
\psi'(x_1)   \end{array} \right), ~~ \Lambda_{\varepsilon lr}= 
\Lambda_{3,l} \Lambda_{0}\Lambda_{2,l}\Lambda_{0}\Lambda_{1,l}  =: \left(
\begin{array}{cc} \bar{\lambda}_{11}~~ \bar{\lambda}_{12} \\
\bar{\lambda}_{21} ~~\bar{\lambda}_{22} \end{array} \right) ,
\label{8}
\end{eqnarray}
connects the boundary conditions for the wavefunction
$\psi(x)$ and its derivative $\psi'(x)$ at $x=x_1= 0 $ and $x=x_2 =  3l+2r $
($N=3$). 
For the case of the two-delta potential ($N=2$) we set in  potential  (\ref{5}) 
$a_3 =0$, so that the boundary conditions are
 $x_1 =0$ and $x_2 = 2l +r$. The matrix elements in (\ref{8}), denoted by 
overhead bars,  
depend on all the shrinking parameters $\varepsilon, ~l$ and $r$, 
whereas in the 
limit matrix elements, if they exist, the bars are omitted, i.e., we  write
 $\lim_{\varepsilon, l, r \to  0 } \Lambda_{\varepsilon l r}  = : \Lambda =
\left( \begin{array}{cc} {\lambda}_{11}~~ {\lambda}_{12} \\
{\lambda}_{21} ~~{\lambda}_{22} \end{array} \right)$.
Having  accomplished the limit procedure,  we set $x_1 =-0$ and 
$\lim_{l,r \to 0}x_2 = +0$.

We follow the notations and the classification of one-point interactions
given by Brasche and Nizhnik \cite{bn}. Thus, we denote
\begin{equation}
\!\!\!\!\!\!\!\!\!\!\!\!\!\!\!\!\!\!\!\!\!\!\!\!\!
\left. \begin{array}{ll}
\psi_s (0) := \psi(+0)- \psi(-0),~~ &\psi'_s (0) := \psi'(+0)- \psi'(-0), \\
\psi_r (0) := \eta \psi(+0)+ (1-\eta) \psi(-0),~~ &
\psi'_r (0) := \eta \psi'(-0)+ (1-\eta) \psi'(+0),
\end{array} \right.
\label{a1}
\end{equation}
where  $\eta \in \R$ is an arbitrary parameter (this is a  
  generalization of the generally accepted case with  $\eta =1/2$,
see, e.g., \cite{g,k,bn,l2}). Then the $\delta$-interaction, or
$\delta$-potential, with intensity $\alpha$ is defined by the boundary conditions
$\psi_s(0)=0$ and $\psi'_s(0)=\alpha \psi_r(0)$, so that
 the $\Lambda$-matrix in this case has the form
\begin{eqnarray}
\Lambda =  \left(
\begin{array}{cc} 1~ ~~~ 0 \\
 \alpha ~~~~ 1 \end {array} \right).
\label{a2}
\end{eqnarray}
The dual interaction is termed a $\delta'$-interaction (the notation has been suggested in 
\cite{a-h,gh87} and adopted in the literature). This point
interaction with intensity $\beta$  defined by the boundary conditions 
$\psi'_s(0)=0$ and $\psi_s(0)=\beta \psi'_r(0)$ has 
 the $\Lambda$-matrix in the form
\begin{eqnarray}
\Lambda =  \left(
\begin{array}{cc} 1~ ~~~ \beta \\
 0 ~~~~ 1 \end {array} \right).
\label{a3}
\end{eqnarray}

As follows from  formulae (\ref{a2}) and (\ref{a3}), 
the usage of the parameter $\eta$ for both the 
$\delta$- and $\delta'$-interactions does not play any role. However, 
for the $\delta'$-potential with intensity $\gamma$ 
the potential part in equation (\ref{1}) is given by 
$\gamma \delta'(x)\psi(x)$ where the wavefunction $\psi(x)$ must be  discontinuous at $x=0$.
Therefore, due to the ambiguity of the product $\delta'(x)\psi(x)$, 
one can suppose the following generalized (asymmetric) averaging in the form
\begin{equation}
\!\!\!\!\!\!\!\!\!\!\!\!\!\!\!\!\!\!\!\!\!\!\!\!\!\!\!
\delta'(x)\psi(x) = \left[ (1-\eta)\psi(-0) +\eta \psi(+0)\right]\delta'(x)+
  \left[ \eta \psi'(-0) +(1-\eta) \psi'(+0)\right] \delta(x) .
\label{b1}
\end{equation}
This suggestion is also motivated by the studies 
 \cite{gw,vs,cnt12} which demonstrate that the plausible averaging with $\eta =1/2$
at the point of singularity in general does not work. 
The $\delta'$-potential with intensity $\gamma$ is defined by the boundary conditions
$ \psi_s(0)=\gamma \psi_r(0)$ and $\psi'_s(0)=- \gamma \psi'_r(0)$ \cite{bn}. 
An equivalent form of these conditions  is given by the $\Lambda$-matrix in the
diagonal form
\begin{eqnarray}
\Lambda =  \left(
\begin{array}{cc} \theta~ ~~~ 0 ~~\\
 0 ~~ ~\theta^{-1} \end {array} \right)
\label{a4}
\end{eqnarray}
with
\begin{equation}
\theta = {1 +(1-\eta)\gamma \over 1-\eta \gamma}.
\label{b2}
\end{equation}

 Finally, instead of the fourth type of point interactions 
 defined in \cite{bn} as $\delta$-magnetic potentials,
in this paper we shall be dealing with potentials which at some (resonant) 
values of intensities
are fully transparent, whereas outside these values they are completely opaque
satisfying the Dirichlet boundary conditions $\psi(\pm 0)=0$. At the resonance sets
the boundary conditions are given by 
the unit matrix $\Lambda =I :=\left( \begin{array}{cc} 1~~ 0 \\
 0 ~~1 \end {array} \right) $. Beside these, as a particular case, 
 resonant-tunnelling $\delta$-potentials will also be shown to exist.

The convergence of the transmission matrix $\Lambda_{\varepsilon, l,r}$ as 
$\varepsilon, l,r \to 0$  can be parameterized through the parameter 
$\varepsilon$ using the powers $\mu>1$ and $\tau>0$ (keeping the same notation used
in \cite{zz11}) as follows 
\begin{equation}
l=\varepsilon^{\mu-1}~~~ \mbox{and}~~~
r = \varepsilon^\tau.
\label{9}
\end{equation}
Then, according to (\ref{7}), we have the following asymptotic relations:
\begin{equation}
k_j \to \sqrt{a_j/ \varepsilon l} = \sqrt{a_j} \,
\varepsilon^{-\mu /2}, ~~
 k_jl \to  \sqrt{a_j} \, \varepsilon^{\mu/2 -1}, ~~
k_j^2 l \to  a_j \varepsilon^{-1}.
\label{10}
\end{equation}
Explicitly, using that $k_j \to \infty$,  $k_j l$ and $k_i/k_j$, $i,j=1,2,3$, 
are finite  and $r \to 0$ as $\varepsilon \to 0$, we find 
the asymptotic behaviour of the elements of the matrix
$\Lambda_{\varepsilon lr } =  \Lambda_{2,l} \Lambda_0 \Lambda_{1,l}$ ($N=2$):
\begin{eqnarray}
\bar{\lambda}_{11} &\to & \cos(k_1 l) \cos(k_2 l) -(k_1/k_2)\sin(k_1 l) \sin(k_2 l)
\nonumber \\
&& - k_1 r \sin(k_1 l) \cos(k_2 l), 
\label{11}\\
\bar{\lambda}_{12}& \to \,\, & 0, \label{12} \\
\bar{\lambda}_{21} &\to & -\, k_1\sin(k_1l)\cos(k_2l) -\, k_2 \cos(k_1l)\sin(k_2l)
\nonumber \\&& + k_1 k_2 r\sin(k_1l)\sin(k_2l), 
\label{13}\\
\bar{\lambda}_{22} &\to & \cos(k_1 l) \cos(k_2 l) -(k_2/k_1)\sin(k_1 l) \sin(k_2 l)
\nonumber \\ && - k_2 r \cos(k_1 l) \sin(k_2 l).
\label{14}
\end{eqnarray}
Similarly, for the three-delta potential the $\bar{\lambda}_{ij}$-asymptotes of
the matrix product $\Lambda_{\varepsilon lr } = \Lambda_{3,l}\Lambda_0 
\Lambda_{2,l} \Lambda_0 \Lambda_{1,l}$ are as follows
\begin{equation*}
\!\!\!\!\!\!\!\!\!\!\!\!\!\!\!\!\!\!\!\!\!\!\!\!\!\!\!\!\!\!\!\!\!\!\!
\bar{\lambda}_{11} 
 \to  \, \cos(k_1 l) \cos(k_2 l)\cos(k_3 l) -(k_1/k_2)\sin(k_1 l) \sin(k_2 l)\cos(k_3 l)
\end{equation*}
\begin{equation*}
\!\!\!\!\!\!\!\!\!\!\!\!\!\!\!\!\!\!
 -\, \, (k_1/k_3)\sin(k_1 l) \cos(k_2 l)\sin(k_3 l)
-(k_2/k_3)\cos(k_1 l) \sin(k_2 l)\sin(k_3 l)
\end{equation*}
\begin{equation*}
\!\!\!\!\!\!\!\!\!\!\!\!\!\!\!\!\!\!
 -\,\,  2k_1 r \sin(k_1 l) \cos(k_2 l)\cos(k_3 l) -k_2 r \cos(k_1 l) \sin(k_2 l)\cos(k_3 l) 
\end{equation*}
\begin{equation}
\!\!\!\!\!\!\!\!\!\!\!\!\!\!\!\!\!\!
  + \,\,  k_1 k_2 r^2 \sin(k_1 l)  \sin(k_2 l)\cos(k_3 l)
 +  (k_1 k_2 r/ k_3)\sin(k_1 l)  \sin(k_2 l)\sin(k_3 l),
\label{15} 
\end{equation}
\begin{equation}
\!\!\!\!\!\!\!\!\!\!\!\!\!\!\!\!\!\!\!\!\!\!\!\!\!\!\!\!\!\!\!\!\!\!\!
\bar{\lambda}_{12}\to \, - \, \, k_2 r^2 \cos(k_1 l) \sin(k_2 l)\cos(k_3 l)  , 
\label{16} 
\end{equation}
\begin{equation*}
\!\!\!\!\!\!\!\!\!\!\!\!\!\!\!\!\!\!\!\!\!\!\!\!\!\!\!\!\!\!\!\!\!\!\!
\bar{\lambda}_{21} \to  \, -\,\,  k_1  \sin(k_1 l) \cos(k_2 l)\cos(k_3 l)
-\, k_2 \cos(k_1 l)\sin(k_2 l)\cos(k_3 l) 
\end{equation*}
\begin{equation*}
\!\!\!\!\!\!\!\!\!\!\!\!\!\!\!\!\!\!
 -\, \, k_3 \cos(k_1 l)\cos(k_2 l)\sin(k_3 l)
+ k_1 k_2 r \sin(k_1 l)\sin(k_2 l)\cos(k_3 l) 
\end{equation*}
\begin{equation*}
\!\!\!\!\!\!\!\!\!\!\!\!\!\!\!\!\!\!
 + \, \, 2  k_1 k_3 r \sin(k_1 l)\cos(k_2 l)\sin(k_3 l) 
+ k_2 k_3 r \cos(k_1 l)\sin(k_2 l)\sin(k_3 l) 
\end{equation*}
\begin{equation*}
\!\!\!\!\!\!\!\!\!\!\!\!\!\!\!\!\!\!
+\, \,  k_1 k_3 (k_2^{-1} - k_2 r^2) \sin(k_1 l)\sin(k_2 l)\sin(k_3 l)
\end{equation*}
\begin{equation}
\!\!\!\!\!\!\!\!\!\!\!\!\!\!\!\!\!\!
 + \, \,   k^2 r^2 \cos(k_2 l) [ k_1 \sin(k_1 l)\cos(k_3 l) 
+  k_3  \cos(k_1 l)\sin(k_3 l)],
\label{17} 
\end{equation}
\begin{equation*}
\!\!\!\!\!\!\!\!\!\!\!\!\!\!\!\!\!\!\!\!\!\!\!\!\!\!\!\!\!\!\!\!\!\!\!
\bar{\lambda}_{22} \to \,  \cos(k_1 l) \cos(k_2 l)\cos(k_3 l)  
 - (k_2/k_1) \sin(k_1 l)  \sin(k_2 l)\cos(k_3 l) 
\end{equation*}
\begin{equation*}
\!\!\!\!\!\!\!\!\!\!\!\!\!\!\!\!\!\!
-\,\, (k_3/k_1) \sin(k_1 l)  \cos(k_2 l)\sin(k_3 l)
- (k_3/k_2)\cos(k_1 l) \sin(k_2 l)\sin(k_3l)  
\end{equation*}
\begin{equation*}
\!\!\!\!\!\!\!\!\!\!\!\!\!\!\!\!\!\!
 -2 k_3 r \cos(k_1 l) \cos(k_2 l)\sin(k_3l) 
- k_2 r\cos(k_1 l) \sin(k_2 l)\cos(k_3l) 
\end{equation*}
\begin{equation}
\!\!\!\!\!\!\!\!\!\!\!\!\!\!\!\!\!\!
 +\,\,  k_2k_3 r^2\cos(k_1 l) \sin(k_2 l)\sin(k_3l)
+ (k_2 k_3 r/k_1) \sin(k_1 l)  \sin(k_2 l)\sin(k_3 l).
\label{18}
\end{equation}

It follows from asymptotes  (\ref{10})-(\ref{18}) 
that the cases with $\mu >2$ ($k_jl \to 0$) and $\mu =2$ (as $k_jl$ tends to a non-zero
constant) should be analysed separately. As shown below, 
the analysis of the convergence of the 
corresponding transmission matrices as $\varepsilon \to 0$ leads to quite different results.

\section{Realizing point interactions under the convergence of the 
$\Lambda_{\varepsilon l r}$-matrix along the families of paths with $2 < \mu \le \infty$ 
and $1 \le \tau \le \infty$   } 
	Using (\ref{10}) with $\mu >2$, for all positive $\tau$ we obtain that in the limit 
as $\varepsilon \to 0$ asymptotic relations (\ref{11}), (\ref{13})-(\ref{15}), (\ref{17}) and
(\ref{18}) are reduced to
\begin{eqnarray}
\bar{\lambda}_{11} &\to & \, 1-  a_1 \left( \varepsilon^{\mu -2} + \varepsilon^{\tau - 1 } 
\right)\!,~~~~
\bar{\lambda}_{22} \to 1-  a_2 \left(  \varepsilon^{\mu -2} + \varepsilon^{\tau -1}\right)\!,
 \label{19} \\
\bar{\lambda}_{21} &\to & - (a_1 + a_2)\varepsilon^{-1}   +  
  a_1 a_2 \varepsilon^{\tau -2},~~~N=2;
 \label{20} 
\end{eqnarray}
\begin{equation}
\!\!\!
\left. \begin{array}{ll}
\bar{\lambda}_{11}&  \to ~ 1- \left[( 2 a_1 +a_2 )\varepsilon^{ -1} - 
   a_1 a_2  \varepsilon^{ \tau -2} \right] \left(\varepsilon^{\mu-1} +\varepsilon^\tau \right)\!,\\
\bar{\lambda}_{22}& \to ~
 1- \left[ (  a_2  +2a_3)\varepsilon^{ -1} -  a_2 a_3 \varepsilon^{ \tau -2}
 \right] \left(\varepsilon^{\mu-1} + \varepsilon^\tau \right)\!,
\end{array} \right.
\label{21}
\end{equation}
\begin{eqnarray}
\bar{\lambda}_{21} \to &-& (a_1 + a_2  + a_3)\varepsilon^{-1}  + 
(  a_1 a_2 + 2 a_1 a_3 + a_2 a_3) \varepsilon^{\tau -2} 
+ a_1a_3 \varepsilon^{\mu -3} \nonumber \\
&-& a_1 a_2 a_3 \varepsilon^{2\tau -3}\! +k^2 (a_1 +a_3)
\varepsilon^{2\tau-1},~~~N=3.
 \label{22} 
\end{eqnarray}
In the limit as $\varepsilon \to 0$ both the limit matrix elements $\lambda_{11}$ and
$\lambda_{22}$ must be finite and therefore, as follows from asymptotes (\ref{19}) and
(\ref{21}), the interval $0< \tau < 1$ is not suitable for the existence of point 
interactions. Consequently, the interval $1 \le \tau \le \infty$ has to be considered 
for the further analysis of the convergence as $\varepsilon \to 0$. Then, similarly to
limit  (\ref{12}), from  (\ref{16}) we also have
$\bar{\lambda}_{12} \to -\, a_2 \varepsilon^{2\tau -1} \to 0 $ in the case with $N=3$.
However, the elements $\bar{\lambda}_{21}$ given by asymptotes (\ref{20}) and (\ref{22})
are always divergent as $\varepsilon \to 0$. The only possibility to make these terms finite
is a cancellation of divergences in the shrinking limit. To accomplish such a cancellation
procedure, in virtue of the form of formulae (\ref{19})-(\ref{22}), 
we split the  interval 
$2 \le \mu \le \infty$ into the four sets: $\{2\}$, $(2,3)$, $\{3\}$, $(3, \infty]$ 
and the interval $1 \le \tau \le \infty$ 
into the  sets: $\{1\}$, $(1,2)$, $\{2\}$, $(2, \infty]$.

\begin{figure}
\centerline{\includegraphics[width=1.0\textwidth]{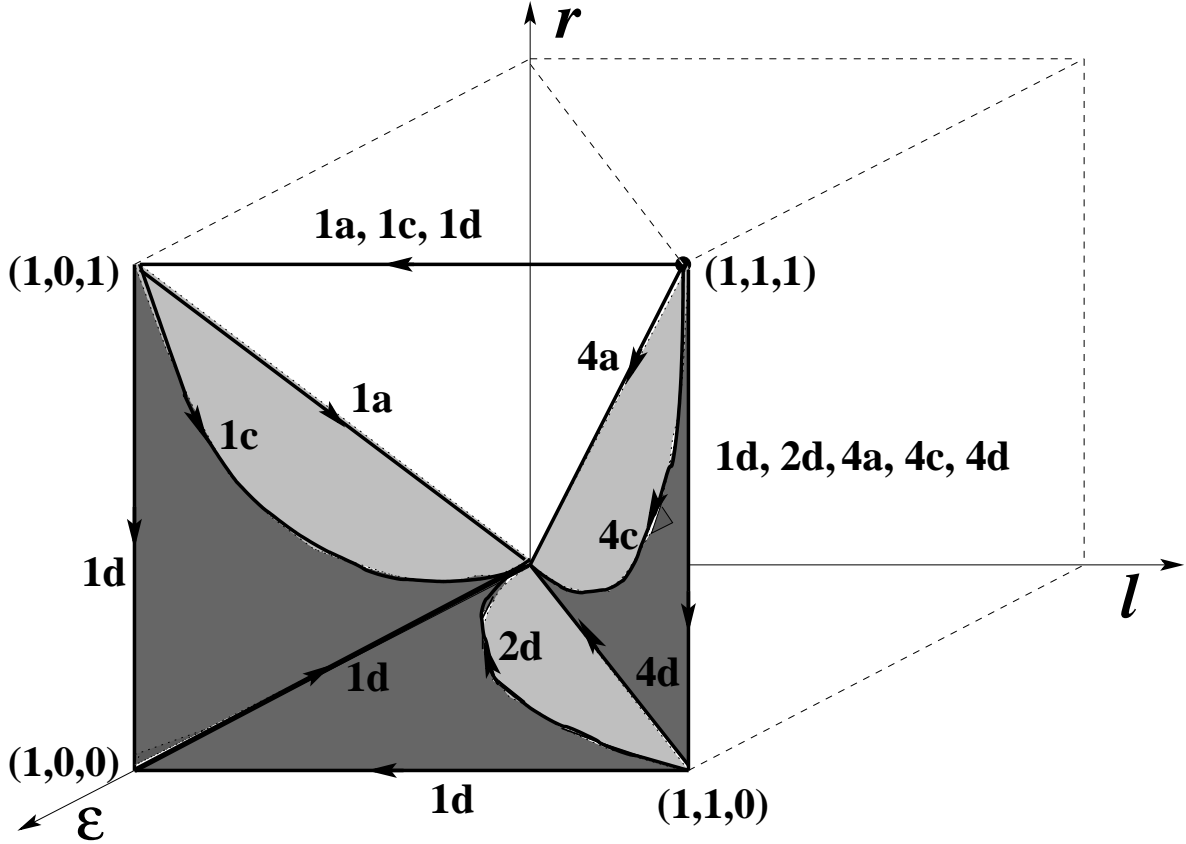}}
\caption{The $(\varepsilon, l,r)$-cube where the eight paths:
 1a ($\mu \to \infty$, $\tau =1$),
 1c ($\mu \to \infty$, $\tau =2$),  2d ($\mu =3$, $\tau \to \infty$), 4a ($\mu = 2,$ $\tau =1$),
4c ($\mu  =\tau =2$), 4d ($\mu  =2$, $\tau \to \infty$) 
and both 1d (first $\mu \to \infty$, then
$\tau \to \infty$ and first $\tau \to \infty$, then
$\mu \to \infty$) are schematically shown by solid lines accompanied with arrows. 
The families of
paths 1b ($\mu \to \infty$, $1< \tau <2$),  2d ($ 2< \mu < 3$, $\tau \to \infty$) and
4b ($\mu =2$, $ 1< \tau < 2$) lie in  sparsely shadowed regions, whereas
the families of paths 1d ($\mu \to \infty$, $2< \tau < \infty$), 
4d ($\mu =2$, $2< \tau < \infty$) and 1d ($3 < \mu < \infty$, $ \tau \to \infty$) 
are illustrated by densely shadowed areas.
 }
\label{fig1}
\end{figure}

Next, for convenience we introduce a three-dimensional system of coordinates 
$(\varepsilon, l, r)$ with the origin at 
$(\varepsilon, l, r) = (0,0,0) = : \{0\}$ and  consider the cube 
  with the vertices at 
$\{0\}$, $(1,0,0)$, $(1,1,0)$, $(0,1,0)$ in the face $r=0$ and 
 $(0,0,1)$, $(1,0,1)$, $(1,1,1)$, $(0,1,1)$ in the face $r=1$, 
as shown in figure~\ref{fig1}.  Then the squeezing limit 
of  potential (\ref{5}) corresponds to 
a path (descent) for which $(\varepsilon, l, r)= (1,1,1)$ is a starting point 
and the origin  $(\varepsilon, l, r) = \{0\}$ a final point. For
the whole interval $1 \le \tau \le \infty$ we consider the  four
families of paths parameterized by $3< \mu \le \infty$ 
(paths 1),  $  \mu =3$ (paths 2), $ 2< \mu <3$ (paths 3) and 
 $ \mu =2$ (paths 4).  In its turn, for the $j$th ($j= \overline{1,4}$) 
family, we also single out the same four subsets: $ja$ ($\tau =1$),   
$jb$ ($ 1< \tau < 2$), $jc$ ($\tau =2$) and $jd$ ($2< \tau  \le \infty$). 
Some of these paths  shown in the faces of the cube and along its edges
are schematically depicted in figure~\ref{fig1}. As shown in this figure, 
in the limit case as $\mu \to \infty$ ($l \to 0$), all the paths of family 1 
follow first  along the edge $(\varepsilon,r)=(1,1)$ and then each of 
these paths descends in the face $l = 0$ approaching the cube origin $\{0\}$
with different rates depending on $\tau$.
Similarly, the  case with $\tau = \infty$ ($r \to 0$) describes the situation
when the squeezing   limit sequentially proceeds along the edge
  $(\varepsilon,l)=(1,1)$ and then along the curve $l=\varepsilon^{\mu-1}$  
in the face $r=0$ with rates depending on $\mu$. Finally, note that
the limit paths when both $\mu$ and $\tau$ tend to infinity are different
depending on the repeated limit: first $\mu \to \infty$, then $\tau \to \infty$ 
or vice versa, first $\tau \to \infty$ and then $\mu \to \infty$.
Below we analyse both connected and separated point interactions which can be 
realized along   all of these
paths   starting at  the point $(1,1,1)$ and ending at the origin  $\{0\} $.

\subsection{Families of paths 1a, 2a and 3a ($1< \mu \le \infty$, $\tau =1$)}

First we note that the $\varepsilon \to 0$ limit of asymptotes (\ref{19}) and (\ref{21})
with $\mu >2$ and $\tau =1$ is finite and therefore this fact ensures
 the existence of point (connected or separated) interactions. Next, as it can be seen
from asymptotes (\ref{20}) and (\ref{22}), for all
 $\mu> 1$ and $\tau>0$ the $\bar{\lambda}_{21}$-terms are divergent as $\varepsilon \to 0$
in both the cases $N=2$ and 3.
The necessary condition to make these terms finite in the
$\varepsilon \to 0$ limit is to impose the equations
\begin{equation}
\!\!\!\!\!\!\!\!\!\!\!\!\!\!\!\!\!\!\!\!\!\!\!\!\!\!\!\!\!\!\!\!\!\!\!\!\!\!
\left. \begin{array}{ll}
 ~~~~ K_2(a_1,a_2) : = a_1 +a_2 -a_1a_2 =0 ~~ & \mbox{for}~~  N=2,  \\
  K_3 (a_1,a_2,a_3) : =  a_1 +a_2 + a_3 -a_1 a_2 -2  a_1 a_3 -
  a_2 a_3 + a_1 a_2 a_3 =0~~ & \mbox{for}~~  N=3. 
\end{array} \right.
 \label{23}
\end{equation}
The first of these equations obtained earlier by Brasche and Nizhnik \cite{bn}
ensures the finiteness of the limit term
 $\lambda_{21}$ for all  paths 1a, 2a and 3a, while for $N=3$, because of the 
presence of the term with $\varepsilon^{\mu -3}$ in (\ref{22}), the existence of connected
interactions is impossible for path 3a (in virtue of the inequality $\mu < 3$). 
Using  equations (\ref{23}) in asymptotes (\ref{19}) and (\ref{21}), we obtain 
the diagonal elements of the limit matrix $\Lambda$ (except for paths 3a with $N=3$):
\begin{equation}
\left. \begin{array}{ll}
\lim_{\varepsilon \to 0}\bar{\lambda}_{11}  =:
\theta = \left\{ \begin{array}{ll}
 1-a_1 &   \mbox{for}~~  N=2, \\
1 - 2a_1 - a_2 +a_1  a_2  & \mbox{for} ~~ N=3,
\end{array} \right.  \\
\lim_{\varepsilon \to 0}\bar{\lambda}_{22} 
 =: \rho = \left\{ \begin{array}{ll}
 1-a_2 &   \mbox{for}~~  N=2, \\
1 - a_2 - 2a_3 +a_2  a_3  & \mbox{for} ~~ N=3.
\end{array} \right. \end{array} \right.
\label{24}
\end{equation}
In virtue of equations (\ref{23}), we have $\rho =\theta^{-1}$ and therefore
for paths 1a, 2a, 3a ($N=2$) and 1a ($N=3$)
 the limit transmission matrix becomes of diagonal form (\ref{a4}). Due to (\ref{b2}), this 
  occurs at the following values of $(a_1,a_2)$ and $(a_1,a_2,a_3)$:
\begin{equation}
a_1 = - {\gamma \over 1- \eta \gamma },~~
a_2 = {\gamma \over 1+ (1-\eta) \gamma}
\label{26}
\end{equation}
for $N=2$ and 
\begin{equation}
a_1 = {1 \over a_2 -2} \left( a_2 + { \gamma \over 1 - \eta \gamma  } \right), ~~
a_3 = {1 \over a_2 -2} \left( a_2 - {\gamma \over 1 + (1 - \eta) \gamma }\right), 
\label{26a}
\end{equation}
with arbitrary $a_2 \in \R \setminus \{2\}$,  for $N=3$.
 We call those intensities $(a_1,a_2) \in \R^2 \setminus \{0\}$ and  
$(a_1,a_2,a_3) \in \R^3 \setminus \{0\}$ which satisfy equations
(\ref{23}) the resonance sets ${\cal K}_2$ and ${\cal K}_3$, respectively.
For $N=2$ the first equation (\ref{23}) describes a curve on the $(a_1,a_2)$-plane, whereas
for $N=3$ we have a surface in the $(a_1,a_2,a_3)$-space. Therefore
the point interactions realized on the sets given by equations (\ref{23}) 
along paths 1a, 2a, 3a ($N=2$) and 1a ($N=3$) may be 
called `single-resonant-tunnelling $\delta'$-potentials of the ${\cal K}$-type'. 

There exists a particular  subfamily  of
the intensities $(a_1, a_2) \in \R^2 \setminus \{0\}$ and 
$(a_1, a_2,  a_3) \in \R^3 \setminus \{0\}$ from the ${\cal K}_{2,3}$-sets 
 for which $\theta = \pm 1$ in (\ref{a4}), realizing
 the point interactions with full transmission.  Thus, for $N=2$ these values are
$a_1 =a_2 =2$ resulting in the unit matrix $\Lambda =-I$. 
In the three-delta case the two conditions $a_1 =a_3$ and $2a_1 +a_2 -a_1a_2  = 0$ 
provide the unit matrix $\Lambda =I$, whereas the other two conditions
$a_1 +a_3 =2$ and $2a_1 +a_2 -a_1a_2 =2$
(in general, an asymmetric structure) lead to the matrix $\Lambda =-I$.

As regards path 2a in the case with $N=3$,  the cancellation of divergences in (\ref{22}) 
leads in the limit as  $\varepsilon \to 0$ to a non-zero constant.
 As a result, in virtue of
(\ref{19}) and (\ref{21}), we have the same limit diagonal elements (\ref{24}) and the 
limit transmission matrix of the form
\begin{eqnarray}
\Lambda =  \left(
\begin{array}{cc} \theta ~~~ 0 ~~\\
\alpha ~~\theta^{-1} \end{array} \right) , ~~~~
\alpha := \lim_{\varepsilon \to 0}\bar{\lambda}_{21},
\label{27}
\end{eqnarray}
with $\alpha = a_1 a_3$.  Therefore,
 to be in agreement with the notation introduced above for paths 
1a, 2a, 3a ($N=2$) and 1a ($N=3$), the 
point interaction realized along path 2a ($\mu=3$, $\tau=1$) on the 
resonance ${\cal K}_3$-set
 may be called   a `single-resonant-tunnelling $(\delta' + \delta)$-potential
of the ${\cal K}$-type'.    

Finally, as follows from asymptote (\ref{22}) for $N=3$,
the cancellation of divergences 
in the limit as $\varepsilon \to 0$ is impossible. Thus, 
everywhere beyond the ${\cal K}_2$-set (paths 1a, 2a, 3a), the 
${\cal K}_3$-set (paths 1a, 2a) and for all $(a_1, a_2, a_3) \in {\R}^3 \setminus \{0\}$,
  the limit point interactions
are separated. They are described by the boundary conditions of the Dirichlet type: 
$\psi(\pm 0)=0 $.

\subsection{Families of paths $jb$, $jc$ and $jd$, $j=1,2,3$}

It follows from asymptotic relations (\ref{19}) and (\ref{21}) that for all  paths 
$jb$, $jc$ and $jd$, $j=1,2,3$, we have the limits
$\bar{\lambda}_{11}, \bar{\lambda}_{22} \to 1$, so that in these cases either 
connected or separated point interactions can be realized. The $\varepsilon \to 0$ analysis has 
to be carried out only for the $\bar{\lambda}_{21}$-terms given by asymptotes 
(\ref{20}) and (\ref{22}). 

{\it Families of paths} 1b, 2b and 3b ($2 < \mu \le \infty$, $1 < \tau < 2$): 
Along these paths the 
$\bar{\lambda}_{21}$-terms are divergent for all non-zero $a_1, a_2$ and $a_3$.
However, there exists a possibility to cancel the divergences in (\ref{22}) 
for paths 1b and 2b ($3 \le \mu \le \infty$) at $\tau =3/2$ and for path 3b at 
$\mu -1 =\tau =3/2$. In the former case the last term in (\ref{22}) is finite and 
the cancellation occurs if both
the equations $a_1 + a_2 + a_3 =0$ and $a_1 a_2 + 2 a_1 a_3 + a_2 a_3 =0$
are fulfilled simultaneously. Similarly, in the latter case, instead of the last
equation, we have  $a_1 a_2 + 3 a_1 a_3 + a_2 a_3 =0$. 
  Excluding $a_3$ from these equations, we find the conditions
$a_1^2 + (a_1 +a_2)^2 =0$ and $(3/4)a_1^2 + (3a_1/2 +a_2)^2 =0$, respectively,
 which are valid only if $a_1 =a_2 =0$ and therefore $a_3 =0$.
Therefore the limit point interactions realized along the family of paths 
1b, 1a and 1c are separated for all non-zero $a_1, a_2$ and $a_3$ with the boundary conditions 
$\psi(\pm 0)=0$.  

{\it Families of paths} 1c, 2c and 3c ($2 < \mu \le \infty$, $ \tau = 2$): 
Contrary to the previous case, for these paths 
 the cancellation of divergences in the $\bar{\lambda}_{21}$-terms is possible,
 except for paths 3c with $N=3$ because of the presence of the 
term with $\varepsilon^{\mu -3}$ in (\ref{22}). As a result, 
a non-zero finite limit of the $ \bar{\lambda}_{21}$-terms
takes place if the conditions 
\begin{equation}
\left. \begin{array}{ll}
   ~~~~ L_2(a_1,a_2)  : =  a_1 +a_2 =0 ~~ & \mbox{for}~~N=2,  \\
   L_3(a_1,a_2,a_3) : =   a_1 +a_2 + a_3=0 ~~ & \mbox{for}~~N=3
\end{array} \right.
\label{29}
\end{equation}
hold true, being just a `linearized' version of equations (\ref{23}). In the following
we refer the intensities $(a_1,a_2) \in \R^2 \setminus \{0\}$ and 
$(a_1,a_2,a_3) \in \R^3 \setminus \{0\}$, which satisfy equations (\ref{29}),
to as ${\cal L}_2$- and ${\cal L}_3$-sets,
respectively. On these sets the limit $\Lambda$-matrix 
describes  the $\delta$-potential with intensity $\alpha$.
From asymptote (\ref{20}) we obtain $ \alpha =  a_1 a_2$ for  $N=2$, while for $N=3$
asymptote (\ref{22}) results in 
\begin{equation}
\alpha=  \left\{ \begin{array}{ll}
   a_1 a_2 +2a_1 a_3 + a_2 a_3 &   \mbox{for paths 1c}, \\
 a_1 a_2 +3a_1 a_3 + a_2 a_3  & \mbox{for path 2c}.
\end{array} \right. 
\label{31}
\end{equation} 
Everywhere beyond the ${\cal L}_2$-set for paths 1c, 2c, 3c and 
  the ${\cal L}_3$-set for paths 1b, 2b as well as for all
$(a_1,a_2,a_3) \in \R^3 \setminus \{0\}$  for paths 3b, the point interactions
are separated satisfying the Dirichlet boundary conditions.

 {\it Families of paths} 1d, 2d and 3d ($2<\mu \le \infty$, $2 < \tau \le \infty$):
The  case with $\tau = \infty$ describes the situation
when the squeezing   limit sequentially follows the edge
  $r \to 0$ and then it goes along the curves $l=\varepsilon^{\mu-1}$ in the face $r=0$
 as shown in figure~\ref{fig1}.  The total cancellation of divergences takes place
for paths 1d, 2d, 3d ($N=2$,  on  the ${\cal L}_2$-set) and 1d 
($N=3$, on the ${\cal L}_3$-set), resulting in the existence of the resonant
point interactions with full transmission ($\Lambda =I$).
In virtue of the term with $\varepsilon^{\mu -3}$ in (\ref{22}), the limit 
point interactions for paths 2d are of the $\delta$-potential type described by
$\Lambda$-matrix (\ref{a2}) with $\alpha =a_1 a_3$. Finally, for paths 3d ($N=3$)
the cancellation of divergences in asymptotes (\ref{22}) as $\varepsilon \to 0$ is
impossible because of the presence of the term with $\varepsilon^{\mu -3}$. Consequently,
for paths 1d, 2d, 3d outside the ${\cal L}_2$-set; 1d, 2d outside the ${\cal L}_3$-set 
and for all $(a_1,a_2,a_3) \in \R \setminus \{0\}$ for paths 3d, the limit interactions
are separated satisfying the Dirichlet boundary conditions.

Thus, the $\Lambda$-matrices for paths 1b, 1c and 1d exhibit  
   the transition of transmission that occurs on the  ${\cal L}_{2,3}$-sets while
varying the rate of the decrease of distance $r$ between the $\delta$-potentials.
For sufficiently slow squeezing this distance ($1 < \tau < 2$, paths 1b), the limit
point interactions  are opaque, for intermediate shrinking 
($\tau =2$, paths 1c) the interactions become  partially 
transparent ($\delta$-potential) and for fast shrinking ($2 < \tau < \infty $)
the interactions appear to be fully transparent. One can check that 
 these results are in agreement with those established by \v{S}eba for $N=2$
 in the limit case $\mu \to \infty$ (see Theorem 3 in  \cite{s}).

\section{Realizing point interactions under the convergence of the 
$\Lambda_{\varepsilon l r}$-matrix along the families of paths with $ \mu =2$ 
and $1 \le \tau \le \infty$   } 

Consider now the situation when $\mu =2$ and $0< \tau \le \infty$. 
Then $l=\varepsilon$ and according to asymptotic relations (\ref{10}), we have 
$k_j \to \sqrt{a_j}/\varepsilon$. 
In this case $r \to 0$ as $\varepsilon \to 0$ and therefore   asymptotes 
 (\ref{11}), (\ref{13}) and (\ref{14})  are reduced to  
\begin{equation}
\!\!\!\!\!\!\!\!\!\!\!\!\!\!\!\!\!\!\!\!\!\!\!\!\!\!\!\!\!\!\!\!\!
\left. \begin{array}{ll}
\bar{\lambda}_{11} \to & 
\cos\!\sqrt{a_1 } \cos\!\sqrt{a_2 }- \sqrt{a_1/a_2 }\sin\!\sqrt{a_1}\sin\!\sqrt{a_2}
 - \sqrt{a_1 } \sin\!\sqrt{a_1 } \cos\!\sqrt{a_2 }\, \varepsilon^{\tau -1},\\
\bar{\lambda}_{22} \to & 
\cos\!\sqrt{a_1 } \cos\!\sqrt{a_2 }- \sqrt{a_2/a_1 }\sin\!\sqrt{a_1}\sin\!\sqrt{a_2}
- \sqrt{a_2 } \cos\!\sqrt{a_1 } \sin\!\sqrt{a_2 }\, \varepsilon^{\tau -1} ,
\end{array} \right.
\label{33}
\end{equation}
\begin{equation}
\!\!\!\!\!\!\!\!\!\!\!\!\!\!\!\!\!\!\!\!\!\!\!\!\!\!\!\!\!\!\!\!\!\!\!\!\!\!
\bar{\lambda}_{21} \to  -(\sqrt{a_1 }\sin\!\sqrt{a_1 }\cos\!\sqrt{a_2 }
+\sqrt{a_2 }\cos\!\sqrt{a_1 }\sin\!\sqrt{a_2 } \, )\varepsilon^{-1}
 + \! \sqrt{a_1 a_2 }\sin\!\sqrt{a_1 }\sin\!\sqrt{a_2 }\, \varepsilon^{\tau-2}
\label{34}
\end{equation}
for $N=2$. Similarly, in the case with $N=3$ asymptotes (\ref{15}), (\ref{17}) and (\ref{18})
are transformed to  
\begin{equation}
\!\!\!\!\!\!\!\!\!\!\!\!\!\!\!\!\!\!\!\!\!\!\!\!\!\!\!\!\!\!\!\!\!\!\!\!\!\!\!\!
\left. \begin{array}{llllllll}
\bar{\lambda}_{11}
 &\to & \cos\!\sqrt{ a_1} \cos\!\sqrt{ a_2}\cos\!\sqrt{a_3} 
 -\sqrt{a_1/a_2} \sin\!\sqrt{ a_1} \sin\!\sqrt{ a_2}\cos\!\sqrt{ a_3}
 \\& &-\, \sqrt{a_1/a_3}\sin\!\sqrt{ a_1} \cos\!\sqrt{ a_2}\sin\!\sqrt{a_3}
 -\sqrt{a_2/a_3} \cos\!\sqrt{ a_1}\sin\!\sqrt{ a_2}\sin\!\sqrt{ a_3}\, 
 \\ && + \left(\! \sqrt{ a_1 a_2/a_3} \sin\!\sqrt{ a_1}  \sin\!\sqrt{a_2}
\sin\!\sqrt{ a_3} - 2  \sqrt{a_1}   \sin\!\sqrt{ a_1} \cos\!\sqrt{ a_2} \cos\!\sqrt{ a_3}\right.
\\&& - \left. \sqrt{a_2}  \cos\!\sqrt{ a_1} \sin\!\sqrt{ a_2} \cos\!\sqrt{a_3} \, \right) \!
\varepsilon^{\tau -1}  
 +  \sqrt{a_1 a_2}  \sin\!\sqrt{a_1} \sin\!\sqrt{ a_2}\cos\!\sqrt{a_3}\, 
\varepsilon^{2(\tau -1)} ,
 \\
\bar{\lambda}_{22} &\to & \cos\!\sqrt{ a_1} \cos\!\sqrt{ a_2}\cos\!\sqrt{ a_3} 
-\sqrt{ a_2/a_1} \sin\!\sqrt{ a_1} \sin\!\sqrt{ a_2}\cos\!\sqrt{ a_3} 
 \\
&&- \, \sqrt{a_3/a_1} \sin\!\sqrt{ a_1}\cos\!\sqrt{ a_2}\sin\!\sqrt{a_3}
 - \sqrt{ a_3/a_2} \cos\!\sqrt{ a_1} \sin\!\sqrt{ a_2}\sin\!\sqrt{ a_3} 
 \\ && + \left(\!\sqrt{ a_2 a_3/a_1} \sin\!\sqrt{ a_1}  \sin\!\sqrt{ a_2}\sin\!\sqrt{ a_3}
-2 \sqrt{a_3}   \cos\!\sqrt{ a_1} \cos\!\sqrt{a_2} \sin\!\sqrt{ a_3}\right.
 \\ && -\left.
\sqrt{a_2}  \cos\!\sqrt{a_1} \sin\!\sqrt{a_2} \cos\!\sqrt{ a_3} \,\right)\! \varepsilon^{\tau -1}  
 +  \sqrt{a_2 a_3}  \cos\!\sqrt{a_1} \sin\!\sqrt{ a_2}\sin\!\sqrt{a_3}\, 
\varepsilon^{2(\tau -1)}, \end{array} \right.
\label{35}
\end{equation}
 \begin{equation*}
\!\!\!\!\!\!\!\!\!\!\!\!\!\!\!\!\!\!\!\!\!\!\!\!\!\!\!\!\!\!\!\!\!\!\!\!\!\!
\left. \begin{array}{lllll}
\bar{\lambda}_{21} \to &  \left(\!
\sqrt{ a_1 a_3/a_2 } \sin\!\sqrt{ a_1}\sin\!\sqrt{ a_2}\sin\!\sqrt{ a_3}
-\sqrt{ a_1}  \sin\!\sqrt{ a_1} \cos\!\sqrt{ a_2}\cos\!\sqrt{ a_3} \right.\\ &
\left. -\, \sqrt{ a_2} \cos\!\sqrt{ a_1}\sin\!\sqrt{ a_2}\cos\!\sqrt{ a_3}  
- \sqrt{ a_3} \cos\!\sqrt{ a_1}\cos\!\sqrt{ a_2}\sin\!\sqrt{ a_3}\, \right)\! \varepsilon^{-1}  \\
 &+ \left(  \sqrt{a_1 a_2}\sin\!\sqrt{ a_1}\sin\!\sqrt{ a_2}\cos\!\sqrt{ a_3} 
+2 \sqrt{a_1 a_3}\sin\!\sqrt{ a_1}\cos\!\sqrt{ a_2}\sin\!\sqrt{ a_3} \right.\\
& + \left. \sqrt{a_2 a_3}\cos\!\sqrt{ a_1}\sin\!\sqrt{ a_2}\sin\!\sqrt{ a_3}\, \right)\!
\varepsilon^{\tau -2} 
- \sqrt{ a_1 a_2 a_3} \sin\!\sqrt{ a_1}\sin\!\sqrt{ a_2}\sin\!\sqrt{ a_3}\, 
\varepsilon^{2\tau -3}
\end{array} \right.
\end{equation*}
\begin{equation}
\!\!\!\!\!\!\!\!\!\!\!\!\!\!\!
+\,\, k^2 \! \cos\!\sqrt{ a_2}\left( \sqrt{ a_1} \sin\!\sqrt{ a_1}\cos\!\sqrt{ a_3}+
\sqrt{ a_3} \cos\!\sqrt{ a_1}\sin\!\sqrt{ a_3} \,\right)\! \varepsilon^{2\tau -1}. 
\label{36} 
\end{equation}

For the realization of (both connected and separated)  interactions in the squeezing limit 
 the elements $\bar{\lambda}_{11}$ and $\bar{\lambda}_{22}$ given by asymptotes (\ref{33})
and (\ref{35}) must be finite as $\varepsilon \to 0$. Consequently, similarly to the case with
$\mu > 1$,  the interval 
$0 < \tau < 1$ is not suitable for realizing point interactions and therefore
we have to consider the region $1 \le \tau \le \infty$. Then 
limit (\ref{16}) becomes $\bar{\lambda}_{12} \to - \sqrt{a_2} \sin\!\sqrt{a_2}\, 
\varepsilon^{2\tau -1} \to 0$.

All the paths of family 4 ($\mu =2$ and $1 \le \tau \le \infty$) are
schematically shown in figure~\ref{fig1}, starting at $(\varepsilon, l,r) =
(1,1,1,)$ and ending at the cube origin $(\varepsilon, l,r)= \{0\}$ 
within the diagonal plane: 4a ($\tau =1$ ), 4b ($1 < \tau < 2$),
4c ($\tau =2$) and 4d ($2< \tau \le \infty$) including the limit $\tau \to \infty$
[first $r \to 0$ along the edge $(\varepsilon, l)=(1,1)$  and then along the 
diagonal $l =\varepsilon \to 0$ in the face $r=0$].

{\it Paths} 4a and 4b ($\mu =2$, $1 \le \tau < 2$): 
The cancellation of divergences in (\ref{34}) at $\tau =1$ 
leads to the resonance equation 
\begin{equation}
\!\!\!\!\!\!\!\!\!\!\!\!\!\!\!\!\!
 F_2(a_1, a_2) : =
 \sqrt{a_1 }\tan\!\sqrt{a_1 } +\sqrt{a_2 }\tan\!\sqrt{a_2 }
-  \sqrt{ a_1 a_2} \tan\!\sqrt{a_1 } \tan\!\sqrt{a_2 } =0 
\label{37}
\end{equation}
for $N=2$. Using this equation in
relations (\ref{33}) at $\tau =1$, we obtain the diagonal limit elements of the 
$\Lambda$-matrix in one of the following forms: 
\begin{equation}
\!\!\!\!\!\!\!\!\!\!\!\!\!\!\!\!
\left. \begin{array}{ll}
{\lambda}_{11}& = ( \cos\!\sqrt{a_1} - \sqrt{a_1}\sin\!\sqrt{a_1} \, )/
 \cos\!\sqrt{a_2} =- \sqrt{a_1}\sin\!\sqrt{a_1} / \sqrt{a_2}\sin\!\sqrt{a_2}\, ,\\
{\lambda}_{22}& = ( \cos\!\sqrt{a_2} - \sqrt{a_2}\sin\!\sqrt{a_2} \,)/
 \cos\!\sqrt{a_1}= - \sqrt{a_2}\sin\!\sqrt{a_2} / \sqrt{a_1}\sin\!\sqrt{a_1}\,.
\end{array} \right.
\label{38}
\end{equation}
Using again  equation (\ref{37}), one can check that the equality 
$\lambda_{11}\lambda_{22} =1$ holds true for matrix elements (\ref{38}). 
Similarly, for the three-delta case the cancellation of divergences in (\ref{36})
at $\tau =1$ results in the resonance equation  
\begin{equation*}
\!\!\!\!\!\!\!\!\!\!\!\!\!\!\!\!\!\!\!\!\!\!\!\!\!\!\!\!\!\!\!\!\!\!\!
F_3(a_1, a_2, a_3) : =
\sum_{j=1}^3 \sqrt{ a_j}  \tan\!\sqrt{ a_j} - 
\sqrt{a_1 a_2}\tan\!\sqrt{ a_1}\tan\!\sqrt{ a_2} 
- 2\sqrt{a_1 a_3}\tan\!\sqrt{ a_1}\tan\!\sqrt{ a_3}
\end{equation*}
\begin{equation}
 -\, \sqrt{a_2 a_3}\tan\!\sqrt{ a_2}\tan\!\sqrt{ a_3}
 + ( a_2 -1) \sqrt{ a_1 a_3 \over a_2}  \prod_{j=1}^3 \tan\!\sqrt{ a_j} =0 .
\label{39}
\end{equation}
Using equation (\ref{39}) in asymptotic relations (\ref{35}) at $\tau =1$, 
the expressions for the limit elements $\lambda_{11}$ and $\lambda_{22}$ 
can be simplified. As a result, we obtain the most simple representation of these
elements:
\begin{equation}
\!\!\!\!\!\!\!\!\!\!\!\!\!\!\!\!\!\!\!\!\!\!\!\!\!\!\!\!\!
\left. \begin{array}{llllll}
\lambda_{11}& = [ \cos\!\sqrt{a_1 }\cos\!\sqrt{a_2 } -2 \sqrt{a_1}\sin\!\sqrt{a_1 }
\cos\!\sqrt{a_2 } -\sqrt{a_2}\cos\!\sqrt{a_1 }\sin\!\sqrt{a_2 } \\
&+ ( \sqrt{a_1a_2}-\sqrt{a_1/a_2} \, ) \sin\!\sqrt{a_1 }\sin\!\sqrt{a_2 }   \, ]/ 
\cos\!\sqrt{a_3 } =
( \sqrt{a_1 a_2 } \sin\!\sqrt{ a_1}\sin\!\sqrt{ a_2} \\
& -\sqrt{ a_1} \sin\!\sqrt{ a_1}\cos\!\sqrt{ a_2} -
  \sqrt{ a_2} \cos\!\sqrt{ a_1}\sin\!\sqrt{ a_2}\, )/\sqrt{ a_3} \sin\!\sqrt{ a_3}\, ,\\
\lambda_{22}& = [ \cos\!\sqrt{a_2 }\cos\!\sqrt{a_3 } -2 \sqrt{a_3}\cos\!\sqrt{a_2 }
\sin\!\sqrt{a_3 } - \sqrt{a_2}\sin\!\sqrt{a_2 }\cos\!\sqrt{a_3 } \\
& + ( \sqrt{a_2a_3}-\sqrt{a_3/a_2}\, ) \sin\!\sqrt{a_2 }\sin\!\sqrt{a_3 }  \, ]/ 
\cos\!\sqrt{a_1} = ( \sqrt{a_2 a_3 } \sin\!\sqrt{ a_2}\sin\!\sqrt{ a_3} \\
& - \sqrt{ a_2} \sin\!\sqrt{ a_2}\cos\!\sqrt{ a_3}
 -\sqrt{ a_3} \cos\!\sqrt{ a_2}\sin\!\sqrt{ a_3} \, )/ \sqrt{ a_1} \sin\!\sqrt{ a_1}\, .
\end{array} \right.
\label{40}
\end{equation}
Using again equation (\ref{39}), one can check that the formula $\lambda_{11}\lambda_{22} =1$,
 in which the matrix elements are given by expressions (\ref{40}), holds true. 

The solutions to  transcendental equations (\ref{37}) and (\ref{39}) 
determine countable 
sets of resonance curves on the $(a_1,a_2)$-plane and resonance surfaces
in $(a_1,a_2,a_3)$-space. We refer these resonance curves and surfaces
to as ${\cal F}_2$- and ${\cal F}_3$-sets, respectively.
The limit transmission matrix on these sets is of 
diagonal form (\ref{a4}) with the element $\theta := \lambda_{11} = \lambda_{22}^{-1}$
 given by (\ref{40}), the values of which are determined by the solutions of equations
(\ref{37}) and (\ref{39}).
The  point interactions of this countable family 
may be called  `multiple-resonant-tunnelling
$\delta'$-potentials of the ${\cal F}$-type'.

Next, on the point subsets of ${\cal F}_{2,3}$ defined by
\begin{equation}
\left. \begin{array}{ll}
{\cal P}_2 & : = \{ a_1,a_2 \, \vert \, 
\sin\!\sqrt{a_1}=\sin\!\sqrt{a_2}=0 \}, \\
{\cal P}_3 & : = \{ a_1,a_2, a_3 \, \vert \, 
\sin\!\sqrt{a_1}=\sin\!\sqrt{a_2}=\sin\!\sqrt{a_3}=0 \}
\end{array} \right.
\label{41}
\end{equation}
we have $\Lambda =\pm I$. Note that no symmetry is required
here, i.e., the reflectionless one-point potentials can be realized even if
$a_1 \neq a_2 $ ($N=2$) or $a_1 \neq a_2 \neq a_3$ ($N=3$). 

Concerning paths 4b (both for $N=2$ and 3),  the cancellation of divergences in
the $\bar{\lambda}_{21}$-terms is impossible, except for the ${\cal P}_{2,3}$-subsets
on which  the divergences in (\ref{34}) and (\ref{36}) disappear as well.
 However,  similarly to paths 1b and 2b, 
we have  to analyse the case $\tau =3/2$ in (\ref{36}). Here
 the $\varepsilon \to 0$ limit of $\bar{\lambda}_{21}$ will be finite 
if both the coefficients at $\varepsilon^{-1}$
and $\varepsilon^{\tau -2}$  equal zero  simultaneously resulting in 
two equations. Excluding  from these equations the term
$\sqrt{ a_3}\tan\!\sqrt{ a_3}$, we find the condition 
$ a_1 \tan^2\!\sqrt{ a_1}\cos^2\!\sqrt{ a_2} + \left( \sqrt{ a_1 }\tan\!\sqrt{ a_1}
+\sqrt{ a_2}\tan\!\sqrt{ a_2} \right)^2 =0 $ which cannot be satisfied  
for all $(a_1, a_2, a_3) \in \R^3 \setminus \{0\}$ and therefore the case with $\tau =3/2$
does not produce connected point interactions. Thus, outside the ${\cal F}_{2,3}$-sets
in the case of path 4a and for all $(a_1,a_2) \in \R^2 \setminus {\cal P}_2$ and
$(a_1,a_2,a_3) \in \R^3 \setminus {\cal P}_3$ for paths 4b, the limit interactions 
are  separated satisfying  the boundary conditions $\psi(\pm 0)=0$.

{\it Paths} 4c and 4d ($\mu =2$, $2 \le \tau \le \infty $): 
As follows from asymptotes (\ref{34}) and (\ref{36}) for these paths,
the cancellation of divergences  occurs if the equations 
\begin{equation}
\!\!\!\!\!\!\!\!\!\!\!\!\!\!\!\!\!\!\!\!\!\!\!\!\!\!\!\!\!\!\!
\left. \begin{array}{ll}
~~\,\,\,\, G_2(a_1,a_2) : =
\sqrt{a_1 }\, \tan\!\sqrt{a_1 }+\sqrt{a_2 }\, \tan\!\sqrt{a_2 } =0, & N=2, \\
G_3(a_1,a_2,a_3) : = \sum_{j=1}^3 \sqrt{ a_j}\, \tan \!\sqrt{ a_j}-
 \sqrt{ a_1 a_3/a_2}\, \prod_{j=1}^3\tan \!\sqrt{ a_j}= 0, & N=3,
\end{array} \right.
\label{42}
\end{equation}
 are satisfied.  Using the first of these equations in  (\ref{33})
at $\tau =2$, we find the following two representations for the diagonal elements
of the $\Lambda$-matrix ($N=2$):
\begin{equation}
\lambda_{11}= \lambda_{22}^{-1}=  \cos\!\sqrt{a_1} / \cos\!\sqrt{a_2} = 
-\sqrt{a_1}\sin\!\sqrt{a_1}/\sqrt{a_2}\sin\!\sqrt{a_2}\, .
\label{43}
\end{equation}
Similarly, using the second equation (\ref{42}), we obtain from (\ref{35}) the diagonal 
elements for $N=3$:
\begin{equation}
\!\!\!\!\!\!\!\!\!\!\!\!\!\!\!\!\!\!\!\!\!\!\!
\left. \begin{array}{llll}
{\lambda}_{11}& =  ( \cos\!\sqrt{ a_1}\cos\!\sqrt{ a_2} -
 \sqrt{ a_1/a_2} \sin\!\sqrt{ a_1}\sin\!\sqrt{ a_2} \,)/ \cos\!\sqrt{ a_3} \\
& = - ( \sqrt{ a_1}  \sin\!\sqrt{ a_1}\cos\!\sqrt{ a_2} 
+  \sqrt{ a_2} \cos\!\sqrt{ a_1}\sin\!\sqrt{ a_2 }\, )/ \sqrt{ a_3} \sin\!\sqrt{ a_3} \, ,\\
{\lambda}_{22} &=   ( \cos\!\sqrt{ a_2}\cos\!\sqrt{ a_3} -
 \sqrt{ a_3/a_2} \sin\!\sqrt{ a_2}\sin\!\sqrt{ a_3}\, )/    \cos\!\sqrt{ a_1} \\
& = -( \sqrt{ a_2}  \sin\!\sqrt{ a_2}\cos\!\sqrt{ a_3} 
+  \sqrt{ a_3} \cos\!\sqrt{ a_2}\sin\!\sqrt{ a_3}\, )/ \sqrt{ a_1} \sin\!\sqrt{ a_1} \, .
 \end{array} \right.
\label{44}
\end{equation}
In virtue of the second  equation (\ref{42}), 
 the equality $\lambda_{11}\lambda_{22} =1$, where the elements are given by 
(\ref{44}), holds true. Next,
 as follows from asymptotes (\ref{34}) and (\ref{36}) at $\mu =\tau =2$ (path 4c),
 the off-diagonal elements $\lambda_{21}$ on resonance sets (\ref{42}) 
are in general non-zero. For this path  they are given by
\begin{equation}
\!\!\!\!\!\!\!\!\!\!\!\!\!\!\!\!\!\!\!\!\!\!\!\!\!\!\!\!\!\!\!\!
\lambda_{21}= \left\{ \begin{array}{lll} 
\sqrt{a_1 a_2 }\sin\!\sqrt{a_1 }\sin\!\sqrt{a_2 } \, , & N=2, \\
\sqrt{a_1 a_2}\sin\!\sqrt{ a_1}\sin\!\sqrt{ a_2}\cos\!\sqrt{ a_3} 
+ 2 \sqrt{a_1 a_3}\sin\!\sqrt{ a_1}\cos\!\sqrt{ a_2}\sin\!\sqrt{ a_3} & \\
~~~~~~~~~~~~~~~~~~~~~+\sqrt{a_2 a_3}\cos\!\sqrt{ a_1}\sin\!\sqrt{ a_2}\sin\!\sqrt{ a_3}\, , & N=3 .
\end{array} \right.
\label{45}
\end{equation}
Setting $\theta :=\lambda_{11}=\lambda_{22}^{-1}$ and $\alpha := \lambda_{21}$
where these elements are given by  (\ref{43})-(\ref{45}), we get the family
of point interactions described by the $\Lambda$-matrix of form (\ref{27}).
The elements $\theta$ and $\alpha$ are determined by the countable sets of solutions
to resonance equations (\ref{42}). Therefore, similarly to the point
interaction realized along path 2a on the single-resonance ${\cal K}_3$-set, the last
family may be called `multiple-resonant-tunnelling ($\delta' + \delta$)-potentials of
the ${\cal G}$-type'. 

Some particular cases of the potentials given by equations (\ref{43})-(\ref{45})
should be singled out. 
 First, we note that on the ${\cal P}_{2,3}$-sets,
as follows from  (\ref{43})-(\ref{45}), the limit transmission matrix
$\Lambda = \pm I$. Similarly to (\ref{41}), one can consider the following 
 point subsets of the resonance sets ${\cal G}_{2,3}$:
\begin{eqnarray}
{ \cal Q}_2 \!  \! \! & := & \{  a_1,a_2 \, \vert \, 
\cos\!\sqrt{a_1}=\cos\!\sqrt{a_2}= 0 \}, \nonumber \\
{\cal Q}_3^{(3)} &:=& \{ a_1,a_2, a_3 \, \vert \, \cos\!\sqrt{a_1}
=\cos\!\sqrt{a_2}=\sin\!\sqrt{a_3}=0 \}, \nonumber \\
{\cal Q}_3^{(2)} &:=& \{ a_1,a_2, a_3 \, \vert \, \cos\!\sqrt{a_1}
=\sin\!\sqrt{a_2}=\cos\!\sqrt{a_3}=0 \}, \nonumber \\
{\cal Q}_3^{(1)} &:=& \{ a_1,a_2, a_3 \, \vert \, \sin\!\sqrt{a_1}
=\cos\!\sqrt{a_2}=\cos\!\sqrt{a_3}=0 \} .
\label{46}
\end{eqnarray}
On these subsets matrix elements (\ref{43})-(\ref{45}) are simplified to
\begin{equation}
\theta := \lambda_{11}= \pm \left\{ \begin{array}{ll} 
\sqrt{a_1 /a_2 } \,  & ~\mbox{for}~N=2, \\
\sqrt{a_1 /a_2}\, ,~~  \sqrt{a_1 /a_3} \, , ~~\sqrt{a_2 /a_3} & ~\mbox{for}~N=3 
\end{array} \right.
\label{47}
\end{equation}
and 
\begin{equation}
\alpha := \lambda_{21}= \mp \left\{ \begin{array}{ll} 
\sqrt{a_1 a_2 } \,  & ~\mbox{for}~ N=2, \\
\sqrt{a_1 a_2}\, ,~~  2 \sqrt{a_1 a_3} \, , ~~\sqrt{a_2 a_3} & ~\mbox{for}~N=3 .
\end{array} \right.
\label{48}
\end{equation}
 In the particular case $a_1 =a_2$, $a_1=a_3$ and $a_2= a_3$, 
the multiple-resonant $(\delta' + \delta)$-potentials are reduced to 
the multiple-resonant $\delta$-potentials given by $\Lambda$-matrix (\ref{a2}).

A more general case is the symmetric structure of the regularized
potential for $N=3$ if $a_1 =a_3$ and $a_2$ is arbitrary. 
Here the second resonance condition (\ref{42}) is reduced
to the following two equations:
\begin{equation}
 \sqrt{a_1}\tan\!\sqrt{a_1} = \sqrt{a_2}( \cos\!\sqrt{a_2} \mp 1)/\sin\!\sqrt{a_2}\, .
\label{49}
\end{equation} 
Using these equations in (\ref{44}) and (\ref{45}),  
we get  the limit transmission matrix that describes the two
representations of the $\delta$-potential with
\begin{equation}
\theta = \pm 1~~~\mbox{and}~~~\alpha =\mp 2a_1 \sin^2\!\sqrt{a_1}
\label{50}
\end{equation}
in $\Lambda$-matrix (\ref{27}),
where $a_1$ depends on $a_2$ through equation (\ref{49}).   As regards 
 paths 4d, instead of equation (\ref{45}) we have $\lambda_{21}=0$,
so that for this family of paths $\Lambda = \pm I$. 

Thus, similarly to the families of paths $jb$, $jc$ and $jd$, $j=1,2,3$,
resulting in the one-point interactions with single resonances, 
 the  interactions realized along path 4c ($\tau =2$)
describe an intermediate case with a partial multiple-resonant transmission, while  for paths
4b ($\tau < 2$) the limit interactions are opaque and for paths 4d ($\tau >2$)
they are fully transparent being multiple-resonant as well. 
Everywhere beyond the ${\cal G}_{2,3}$-sets 
we have the separated point interactions with the Dirichlet conditions $\psi(\pm 0)=0$. 
\begin{table}
\caption{\label{tab:table1}
 Resonance sets and transmission matrices for resonant-tunnelling point interactions
realized along all the possible paths in the $(\varepsilon, l, r)$-cube}
\begin{tabular}{|l|l|c|c|c|}
\hline
\hline
Resonant  & Paths & Resonance sets  &    Resonance sets  & $\Lambda$-matrices\\
point interactions &  & ($N=2$) & ($N=3$) & \\
 \hline
$\delta'$-potentials & 1a & ${\cal K}_2 $ & $ {\cal K}_3$ & (\ref{a4}), (\ref{24}) \\
\cline{2-4}
                    & 2a, 3a & ${\cal K}_2$  &   -      &  \\
\cline{2-5}
                    &  4a  & ${\cal F}_2$ & ${\cal F}_3$  & (\ref{a4}), (\ref{38}), (\ref{40}) \\
\cline{2-5}
                    & 4d &   ${\cal G}_2$ & ${\cal G}_3$ & (\ref{a4}), (\ref{43}), (\ref{44}) \\
 \hline
$\delta$-potentials & 1c, 2c & ${\cal L}_2$ & ${\cal L}_3$ & (\ref{a2}), $\alpha =a_1a_2$, (\ref{31})\\
\cline{2-5}
                  & 2d & - &${\cal L}_3$ & (\ref{a2}), $\alpha =a_1a_3$ \\
\cline{2-5}
                    & 3c &$ {\cal L}_2$ & -          & (\ref{a2}), $\alpha =a_1a_2$ \\
\cline{2-5}
         &  4c &  $ {\cal Q}_2,~a_1=a_2$ & - & (\ref{27}), (\ref{47}), (\ref{48})\\
\cline{3-4}
                    &   & - & ${\cal Q}_3^{(3)},~a_1 =a_2$ &   \\
\cline{4-4}
                    &   &  & $ {\cal Q}_3^{(2)},~a_1 =a_3$ &  \\
 \cline{4-4}
                  &   &  & ${\cal Q}_3^{(1)},~ a_2 =a_3$ &  \\
 \cline{4-5} 
               &  &  & $a_1 =a_3,~(\ref{49})$ & (\ref{27}), (\ref{50}) \\
\hline
($\delta' +\delta$)- potentials & 2a & -& $ {\cal K}_3$ & (\ref{24}), (\ref{27}), $\alpha = a_1a_3$ \\
   \cline{2-5}
      & 4c & ${\cal G}_2$ & ${\cal G}_3 $ & (\ref{27}), (\ref{43})-(\ref{45}) \\
\hline
Reflectionless     &  1a      &   - & $a_1 =a_3, $  & $\Lambda =I$  \\
  potentials       &          &     & $2a_1 +a_2 -a_1a_2=0$ &   \\
\cline{3-5}
                   &          &   $a_1=a_2=2$ &  $a_1 +a_3 =2,$ & $\Lambda =-I$ \\
                   &          &              & $2a_1 +a_2 -a_1a_2=2$ &   \\
\cline{2-5}
                   &  1d   & ${\cal L}_2$ & $ {\cal L}_3 $ & $\Lambda =I$ \\
\cline{2-4}
                   & 2d, 3d      & ${\cal L}_2$ & - &      \\
\cline{2-5}
                   & 4a, 4b,  & $  {\cal P}_2$ & ${\cal P}_3$ & $\Lambda = \pm I$  \\
                   & 4c, 4d &&& \\
\cline{2-4}
                    & 4d &   ${\cal Q}_2,~a_1=a_2$ & - & \\
\cline{3-4}
                    &   & - & ${\cal Q}_3^{(3)},~a_1 =a_2$ &   \\
 \cline{4-4}
                    &   &  & ${\cal Q}_3^{(2)},~a_1 =a_3$ &  \\
    \cline{4-4}
                  &   &  & ${\cal Q}_3^{(1)},~a_2 =a_3$ &  \\
   \cline{4-4}
              &&  & $(\ref{49}),~a_1 =a_3$ &  \\
\hline
\hline
\end{tabular}
\end{table}
Note that the
 results given by the first equation (\ref{42}) and formulae (\ref{43}) have been
obtained  earlier in  \cite{c-g,gm}.

\section{Concluding remarks}

The main goal of this paper has been to approximate the system consisting of two and three
$\delta$-potentials (with intensities $a_j \neq 0$, $j=1,2$ if $N=2$ and 
$j=1,2,3$ if $N=3$)
\begin{table}
\caption{\label{tab:table2}
 Non-resonant conditions for separated point interactions realized along all the possible
paths in the $(\varepsilon, l, r)$-cube }
\begin{tabular}{|l|c|c|}
\hline
\hline
 Paths & Non-resonant &  Non-resonant  \\
       & conditions ($N=2$)& conditions ($N=3$) \\
\hline
1a, 2a & $ (a_1,a_2) \notin {\cal K}_2 \cup \{0\} $& $(a_1,a_2,a_3) 
\notin {\cal K}_3 \cup \{0\} $ \\
\hline 
 1b, 2b, 3b & $(a_1,a_2) \in \R^2 \setminus \{0\}$
                          &$(a_1,a_2,a_3) \in \R^3 \setminus \{0\} $\\
                                                       \hline
1c, 1d, 2c, 2d  &$ (a_1,a_2) \notin {\cal L}_2 \cup \{0\} $ & $(a_1,a_2,a_3) 
\notin {\cal L}_3 \cup \{0\}$ \\
\hline
3a & $(a_1,a_2) \notin {\cal K}_2 \cup \{0\}$ & $(a_1,a_2,a_3) \in \R^3 \setminus \{0\} $\\
\hline
3c, 3d & $(a_1,a_2) \notin {\cal L}_2 \cup \{0\}$ & $(a_1,a_2,a_3) \in \R^3 \setminus \{0\} $\\
\hline
4a  & $(a_1,a_2) \notin {\cal F}_2 \cup \{0\}$ & $(a_1,a_2,a_3) \notin {\cal F}_3 \cup \{0\}$ \\
\hline
4b  & $(a_1,a_2) \notin {\cal P}_2\cup \{0\} $ & $(a_1,a_2,a_3) \notin {\cal P}_3 \cup \{0\}$ \\
\hline
4c, 4d & $(a_1,a_2) \notin {\cal G}_2\cup \{0\} $ & $(a_1,a_2,a_3) \notin {\cal G}_3\cup \{0\}$ \\
\hline
\hline
\end{tabular}
\end{table}
by piecewise constant functions and then to investigate the convergence of
the corresponding transmission matrices  in the squeezing limit
as both the width of  $\delta$-like functions $l$ and the
distance between them $r$ tend to zero. The admissible rates of 
shrinking the parameters $l$ and $r$ are controlled through the 
approximation given by equations (\ref{9}), involving the two powers 
$\mu$ and $\tau$ as well as the parameter $\varepsilon \to 0$. For convenience of
the presentation, the three-dimensional $(\varepsilon, l, r)$-cube has  been 
introduced and various paths inside it were considered.
Starting from the {\it same}  three-layer (for $N=2$)
and   five-layer (for $N=3$) potential profile described by (\ref{5}), 
a whole family of  limit one-point interactions 
with resonant-tunnelling behaviour has been realized.  The resonance sets for these 
interactions are curves on the $(a_1,a_2)$-plane  ($N=2$) and surfaces in the 
$(a_1,a_2,a_3)$-space  ($N=3$). The number of resonances (one or infinite) depends on 
a path, along which the corresponding sequence of transmission matrices
has a limit.  

 For both the cases with $N=2$ and 3 we single out
the  four resonance sets named ${\cal K}_{2,3}$, ${\cal L}_{2,3}$,
${\cal F}_{2,3}$, ${\cal G}_{2,3}$ and defined by equations (\ref{23}), (\ref{29}),
(\ref{37}) and (\ref{39}), (\ref{42}), respectively. The first two of these sets
describe {\it single} and  the two others  {\it multiple} resonances.
Accordingly,  the  one-point interactions realized on these sets belong to
 ${\cal K}$-, ${\cal L}$, ${\cal F}$- and ${\cal G}$-families and their $\Lambda$-matrices
are given by (\ref{a2}), (\ref{a4}) and (\ref{27}).
The $\Lambda$-matrix elements for the interactions of the ${\cal K}$- and 
${\cal L}$-families are single-valued, while for  the families ${\cal F}$ and 
${\cal G}$ these elements are multi-valued. All these interactions together with the paths
along which they are realized, including the corresponding resonance sets and 
$\Lambda$-matrices are summarized in table~\ref{tab:table1}.
Here the following  four subfamilies of 
one-point resonant-tunnelling interactions are singled out:
 (i) the  $\delta'$-potentials (single-resonant of the ${\cal K}$- and 
multiple-resonant of the ${\cal F}$-, ${\cal G}$-types),
 (ii) the  $\delta$-potentials (single-resonant of the ${\cal L}$-type, including 
multiple-resonant defined on  the ${\cal Q}$-subsets),
(iii) the  $(\delta' + \delta)$-potentials (single-resonant of the ${\cal K}$- and 
multiple-resonant of the ${\cal G}$-types) and (iv) the reflectionless potentials
(single-resonant of the ${\cal L}$-type, including multiple-resonant defined on  
the ${\cal P}$- and ${\cal Q}$-subsets).

Outside the resonance sets all the one-point interactions become separated with
the Dirichlet boundary conditions $\psi(\pm 0)=0$.   The corresponding conditions
for the existence of this type of interactions given 
on the $(a_1,a_2)$-plane and in the $(a_1,a_2,a_3)$-space depend on the paths and they 
are summarized in table~\ref{tab:table2}.

It should be noticed that in this paper we have 
restricted ourselves to  power parameterization (\ref{9}).   
The admissible set of the powers $\mu$ and $\tau$
for realizing point interactions appears to be  the set
$Q := \{ 2 \le \mu \le \infty \} \times \{ 1 \le \tau \le \infty \}$. 	
Using this parameterization as well as the piecewise constant approximation of 
the $\delta$-functions in potential (\ref{3}), it is possible to get the explicit
solutions for the corresponding $\Lambda$-matrices and to
 treat thus the reflection-transmission properties of the one-point
interactions directly. It is of interest to note that inside the set $Q$ the 
resonance sets are single-valued and when approaching the boundary 
$\{ \mu =2,\, 1 \le \tau \le \infty \}$, the {\it furcation} of the resonance sets occurs.
Qualitatively, all these results are the same for $N=2$ and 3, except for the 
dimension of the resonance sets and the corresponding equations.
 In principle, a similar straightforward 
analysis could be carried out for higher $N$ resulting in 
the same types of one-point interactions with  
resonance sets  ${\cal K}_N$, ${\cal L}_N$, ${\cal F}_N$ and ${\cal G}_N$
being $(N-1)$-dimensional hypersurfaces,
however, the corresponding formulae appear to be quite complicated. 
In the case, if we would like to deal with potentials (\ref{5}) which admit 
distributional limits, for instance, the $\delta'(x)$ potential, 
constraint (\ref{29})  has to be imposed in addition to sets (\ref{23}), 
(\ref{37}), (\ref{39}) and (\ref{42}). Therefore 
this constraint reduces the dimension of resonance sets by one, so that for 
$N=2$, instead of the resonance curves, we have the corresponding set of 
discrete numbers and for $N=3$ 
one-dimensional curves. Some of the particular
cases for $N=2$  have been treated in \cite{c-g,gm,zpla10}. 
 To conclude, it should be noticed that the approach developed in this paper can 
 be a starting point for further studies on regular approximations of point 
interactions and understanding  the resonant mechanism.

\bigskip
{\bf  Acknowledgments}
\bigskip

The financial support from the National Academy of Sciences of Ukraine under project
 No.~0112U000053 is acknowledged.
The author  would like to express gratitude to 
Yaroslav Zolotaryuk for stimulating discussions and valuable suggestions.


\bigskip
{\bf References}
\bigskip 

\begin{thebibliography}{99}

\bibitem{do1}
Demkov~Y~N and Ostrovskii~V~N 1975 Zero-Range Potentials and Their Applications in Atomic Physics
(Leningrad: Leningrad University Press)

\bibitem{do2}
Demkov~Y~N and Ostrovskii~V~N 1988 {\it Zero-Range Potentials and Their Applications in
Atomic Physics} (New York: Plenum)

\bibitem{a-h}
Albeverio~S, Gesztesy~F,  H{\o}egh-Krohn~R and Holden~H 2005 
 {\it Solvable Models in Quantum Mechanics (With an Appendix by Pavel Exner)} 
2nd revised edn  (Providence: RI: American Mathematical Society: Chelsea Publishing)

\bibitem{ak}
Albeverio~S and Kurasov~P 1999 {\it Singular Perturbations of Differential Operators: 
Solvable Schr\"{o}dinger-Type Operators} (Cambridge: Cambridge University Press) 

\bibitem{pc}
Perez~J~F and Coutinho~F~A~B 1991 {\it Am. J. Phys.} {\bf 59} 52

\bibitem{bft}
Brasche~J~F, Figari~R and Teta~A 1998 {\it Potential Analysis}
{\bf 8} 163

\bibitem{enz}
Exner~P, Neidhardt~H and Zagrebnov~V~A 2001
 {\it Commun. Math. Phys.} {\bf 224}  593

\bibitem{c-g}
 Christiansen~P~L, Arnbak~N~C, Zolotaryuk~A~V, Ermakov~V~N and 
Gaididei~Y~B 2003 {\it J. Phys. A: Math. Gen.} {\bf 36} 7589 

\bibitem{ca}
 Coutinho~F~A~B and Amaku~M 2009 {\it Eur. J. Phys.} {\bf 30} 1015

\bibitem{em}
Exner~P and Manko~S~S 2014 {\it Lett. Math. Phys.} {\bf 104} 1079

\bibitem{acf}
Albeverio~S, Cacciapuoti~C and Finco~D 2007 {\it J. Math. Phys.} 
 {\bf 48} 032103

\bibitem{ce} 
 Cacciapuoti~C and Exner~P 2007 {\it J. Phys. A: Math. Theor.}  {\bf 40} F511

\bibitem{tc1}
Turek~O and Cheon~T 2012 {\it Europhys. Lett.}  {\bf 98} 50005 

\bibitem{tc2}
Turek~O and Cheon~T 2013  Ann. Phys. (N.Y.) {\bf 330} 104

\bibitem{z}
Zolotaryuk~A~V 2013 {\it Phys. Rev. A} {\bf 87} 052121 

\bibitem{zz1}
Zolotaryuk~A~V and Zolotaryuk~Y 2015 {\it Phys. Lett. A} 
{\bf 379} 511 

\bibitem{zz2}
Zolotaryuk~A~V and Zolotaryuk~Y 2015 {\it J. Phys. A: Math. Theor.} 
{\bf 48} 035302 

\bibitem{s}
\v{S}eba~P 1986 {\it Rep. Math. Phys.} {\bf 24} 111

\bibitem{gh87} 
 Gesztesy~F and Holden~H 1987 {\it J. Phys. A: Math. Gen.}  {\bf 20} 5157

\bibitem{g}
Griffiths~D~J  1993 {\it J. Phys. A: Math. Gen.} {\bf 26} 2265

\bibitem{k}
Kurasov~P 1996 {\it J. Math. Anal. Appl.} {\bf 201} 297

\bibitem{adk}
Albeverio~S, D\c{a}browski~L and Kurasov~P 1998 {\it Lett. Math. Phys.} {\bf 45} 33

\bibitem{cnp}
Coutinho~F~A~B, Nogami~Y and Perez~J~F 1997 
{\it J. Phys. A: Math. Gen.} {\bf 30} 3937

\bibitem{cnt}
Coutinho~F~A~B, Nogami~Y and Tomio~L 1999 {\it J. Phys. A: Math. Gen.} {\bf 32} 4931

\bibitem{an}
Albeverio~S and Nizhnik~L 2003 {\it Lett. Math. Phys.} {\bf 65} 27

\bibitem{n1}
 Nizhnik~L~N 2003 {\it J. Funct. Anal. Appl.} {\bf 37} 85

\bibitem{n2}
Nizhnik~L~N 2006 {\it J. Funct. Anal. Appl.} {\bf 40} 74

\bibitem{gnn}
Gadella~M, Negro~J and  Nieto~L~M 2009 {\it Phys. Lett.} A  {\bf 373} 1310 

\bibitem{acg}
Arnbak~H, Christiansen~P~L and Gaididei~Y~B 2011 {\it Philos. Trans. R. Soc. A } 
{\bf 369} 1228 

\bibitem{l1}
 Lange~R-J 2012 {\it J. High Energy Phys.} {\bf JHEP11(2012)}, no. 32  

\bibitem{bn}
Brasche~J~F and Nizhnik~L~P 2013 {\it Methods Funct. Anal. Topol.} {\bf 19} 4
(arXiv:1112.2545v1 [math.FA])

\bibitem{gggm}
Gadella~M,  Garc\'{i}a-Ferrero~M~A, 
Gonz\'{a}lez-Mart\'{i}n~S and Maldonado-Villamizar~F~H 2014 
{\it Int. J. Theor. Phys.} {\bf 53} 1614 

\bibitem{l2}
 Lange~R-J 2015 {\it J. Math. Phys.} {\bf 56} 122105

\bibitem{z15}
Zolotaryuk~A~V  2015 {\it J. Phys. A: Math. Theor.} {\bf 48} 255304

\bibitem{kp}
Kulinskii~V~L and Panchenko~D~Y 2015 {\it Physica B: Physics of Condensed Matter} 
{\bf 472} 78

\bibitem{djp}
Dias~N~C, Jorge~C and Prata~J~N 2016 {\it J. Differential Equations} {\bf 260} 6548

\bibitem{gmmn}
Gadella~M, Mateos-Guilarte~J, Mu\~{n}oz-Casta\~{n}eda~J~M and Nieto~L~M 2016 
{\it J. Phys. A: Math. Theor.} {\bf 49} 015204

\bibitem{knt}
Konno~K, Nagasawa~T and Takahashi~R 2016 (arXiv:1605.05418v1 [quant-ph])

\bibitem{km}
Kostenko~A and Malamud~M 2013 {\it Spectral Analysis, Differential Equations and
Mathematical Physics - Proceedings of Symposia in Pure Mathematics} eds H~Holden
{\it et al.} vol. 87 (Providence: RI: American Mathematical Society) p. 235

\bibitem{zci1}
 Zolotaryuk~A~V, Christiansen~P~L and Iermakova~S~V 2006 
{\it J. Phys. A: Math. Gen.} {\bf 39} 9329

\bibitem{tn}
 Toyama~F~M and  Nogami~Y 2007 {\it J. Phys. A: Math. Theor.} {\bf 40} F685 

\bibitem{zz14}
Zolotaryuk~A~V and Zolotaryuk~Y 2014 {\it Int. J. Mod. Phys.} B
{\bf 28} 1350203 

\bibitem{gm}
 Golovaty~Y~D and Man'ko~S~S 2009 {\it Ukrainian Math. Bull.}
 {\bf 6} 169   (e-print arXiv:0909.1034v2 [math.SP])

\bibitem{gh}
Golovaty~Y~D and Hryniv~R~O 2010 {\it J. Phys. A: Math. Theor.}
 {\bf 43} 155204 \\
Golovaty~Y~D and Hryniv~R~O 2011 {\it J. Phys. A: Math. Theor.}
  {\bf 44} 049802  

\bibitem{m1}
Man'ko~S~S 2010 {\it J. Phys. A: Math. Theor.} {\bf 43} 445304

\bibitem{g1}
Golovaty~Y 2012 {\it Methods Funct. Anal. Topol.} {\bf 18} 243

\bibitem{m2}
Man'ko~S~S 2012 {\it J. Math. Phys.} {\bf 53} 123521

\bibitem{gh1}
Golovaty~Y~D and Hryniv~R~O 2013 {\it Proc. R. Soc. Edinb. } A
 {\bf 143} 791   

\bibitem{g2}
Golovaty~Y 2013 {\it Integr. Equ. Oper. Theor.} {\bf 75} 341

\bibitem{zpla10}
Zolotaryuk~A~V 2010 {Phys. Lett. A} {\bf 374} 1636

\bibitem{cs}
Cheon~T and Shigehara~T 1998  {\it Phys. Lett. A} {\bf 243} 111

\bibitem{an1}
Albeverio~S and  Nizhnik~L 2000 {\it Ukr. Mat. Zh.} {\bf 52} 582, 
translation in 2001 {\it Ukr. Math. J.} {\bf 52} 664.

\bibitem{gw}
Griffiths~D and Walborn~S 1999 {\it Am. J. Phys.} {\bf 67} 446

\bibitem{vs}
De Vincenzo~S and Sanchez~C 2010 {\it Can. J. Phys.} {\bf 88} 809

\bibitem{cnt12}
Coutinho~F~A~B, Nogami~Y and Toyama~F~M 2012 {\it Can. J. Phys.} {\bf 90} 383

\bibitem{zz11}
Zolotaryuk~A~V and Zolotaryuk~Y 2011 {\it J. Phys. A: Math. Theor.} 
{\bf 44} 375305 


\end{thebibliography}
\end{document}